\documentclass[12pt]{article}
\usepackage{amsfonts}
\usepackage[latin1]{inputenc}
\usepackage[all]{xy}
\usepackage{graphicx}
\usepackage{latexsym}
\usepackage{color}
\usepackage{epsfig}
\def\l[{\left[}
\def\r]{\right]}

\newcommand{\ti}{\;\;\makebox[0pt]{$\top$}\makebox[0pt]{$\cap$}\;\;}
\def\ba{\begin{array}}
\def\ea{\end{array}}
\def\beq{\begin{equation}}
\def\eeq{\end{equation}}
\def\bea{\begin{eqnarray}}
\def\eea{\end{eqnarray}}

\begin{document}

\pagestyle{empty}
\setcounter{page}{0}
%\hspace{-1cm}\lapth

\begin{center}
{\Large {\bf Deligne-Beilinson cohomology and abelian link invariants: torsion case.}}%
\\[1.5cm]

{\large F. Thuillier}

\end{center}

\vskip 0.7 truecm

{\it LAPTH, Chemin de Bellevue, BP 110, F-74941
Annecy-le-Vieux cedex, France}.

\vspace{20mm}

\centerline{{\bf Abstract}}
%\begin{center}
For the abelian Chern-Simons field  theory,  we  consider the quantum functional integration over the Deligne-Beilinson cohomology classes and present an explicit path-integral non-perturbative computation of the Chern-Simons link invariants in $SO(3)\simeq\mathbb{R}P^3$, a toy example of 3-manifold with torsion.

%\end{center}

\vspace{5mm}

\indent

%\vfill
%\rightline{hep-th-\yymmnnn}
%\rightline{LAPTH-1305/09}
%\newpage
\pagestyle{plain} \renewcommand{\thefootnote}{\arabic{footnote}}

\section{Introduction}

In a quite recent paper \cite{GT}, we have shown how Deligne-Beilinson cohomology \cite{De,Be,EV,Ja,Br,BGST} within Chern-Simons QFT framework \cite{Sc,Ha,Po,Wi,Jo,RT,GMM,Gu} can be used to provide a non perturbative way to compute abelian link invariants on some three dimensional manifolds, such as  $S^3$,  $S^2 \times S^1$ etc. In particular, quantization of the Chern-Simons parameter $k$  as well as the charges $q$  of the links was a straightforward consequence of the use of Deligne-Beilinson cohomology, and the standard regularization via framing was directly interpreted as the problem of regularizing the product of two distributional Deligne-Beilinson cohomology classes.

Actually this former article was only dealing with torsion free (oriented) 3-manifold. We are going to mend this lake of generality by explaining how to extend our approach to (oriented) 3-manifolds with torsion. As a school case, we will consider the oriented 3-manifold $SO(3)\simeq\mathbb{R}P^3$.

In a first section we will recall some basic facts concerning Deligne-Beilinson cohomology and how it relates to the functional measure based on the abelian Chern-Simons action. In a second section we will deal with Wilson lines themselves.
$\\$

\noindent Here are the three results we will obtain:

1) The Chern-Simons level parameter $k$ has to be be even;

2) Trivial cycles give the same result than in $S^3$;

3) Torsion cycles must hold an even charge,

\noindent in perfect agreement with surgery methods.
$\\$

All along this paper we will use the notation $\mathop  = \limits_\mathbb{Z}$, standing for equality modulo $\mathbb{Z}$.

\section{Deligne-Beilinson cohomology: constraints on the level k of the abelian Chern-Simons theory.}

Let us remind that Deligne-Beilinson (DB) cochains can be seen as generalizations of $U(1)$-connections on $U(1)$-principal bundles over smooth manifolds, their classes classifying the corresponding objects, \textit{i.e.} $U(1)$-gerbes with connections \cite{Br,MP}. Concentrating on the case of an oriented 3-manifold $M$, its DB cohomology space $H_D^1 \left({M,{\mathbb Z}} \right)$ is canonically embedded into the following exact sequence \cite{Br,HLZ}:
\begin{equation}
\label{1}
0\buildrel \over \longrightarrow {\Omega ^1\left( M \right)} \mathord{\left/
{\vphantom {{\Omega ^1\left( M \right)} {\Omega _{\mathbb Z}^1 \left( M
\right)}}} \right. \kern-\nulldelimiterspace} {\Omega _{\mathbb Z}^1 \left( M
\right)}\buildrel \over \longrightarrow H_D^1 \left( {M,{\mathbb Z}}
\right)\buildrel \over \longrightarrow \check{H}^{2}\left( {M,{\mathbb Z}}
\right)\buildrel \over \longrightarrow 0 \, ,
\end{equation}
where $\Omega ^1\left( M \right)$ is the space of smooth 1-forms on $M$,
$\Omega _{\mathbb Z}^1 \left( M \right)$ the space of smooth closed 1-forms with
integral periods on $M$ and $\check{H}^{2}\left( {M,{\mathbb Z}} \right)$ is the
second integral $\check{\textrm{C}}$ech cohomology group of $M$. Actually,
$H_D^1 \left({M,{\mathbb Z}} \right)$ can also be embedded into (\cite{HLZ})
\begin{equation}
\label{2}
0\buildrel \over \longrightarrow \check{H}^{1}\left( {M,{\mathbb R
}/{\mathbb Z}}\right)\buildrel
\over \longrightarrow H_D^1 \left( {M,{\mathbb Z}}
\right)\buildrel \over \longrightarrow \Omega _{\mathbb Z}^2 \left( M \right)\buildrel
\over \longrightarrow 0 \, ,
\end{equation}
where $\check{H}^{1}\left( {M,{\mathbb R}/{\mathbb Z}}\right)$ is the first ${\mathbb R
}/{\mathbb Z}$-valued $\check{\textrm{C}}$ech cohomology group of  $M$ and $\Omega _{\mathbb Z}^2 \left( M \right)$ the space of smooth closed 2-forms with integral periods on $M$. Each one of these two exact sequences has its own interest to describe $H_D^1 \left({M,{\mathbb Z}} \right)$, but both give this space the structure of an affine bundle, with (discrete) base $\check{H}^{2}\left( {M,{\mathbb Z}} \right)$  and translation group ${\Omega ^1\left( M \right)} \mathord{\left/
{\vphantom {{\Omega ^1\left( M \right)} {\Omega _{\mathbb Z}^1 \left( M
\right)}}} \right. \kern-\nulldelimiterspace} {\Omega _{\mathbb Z}^1 \left( M
\right)}$  from the former sequence, and with base $\Omega _{\mathbb Z}^2 \left( M \right)$ and translation group $\check{H}^{1}\left( {M,{\mathbb R}/{\mathbb Z}}\right)$ from the latter one.

The other important DB space we will need is $H_D^3 \left({M,{\mathbb Z}} \right)$. However, the exact sequences of the previous type into which this space is embedded both lead to  $H_D^3 \left({M,{\mathbb Z}} \right)\simeq{{\mathbb R}/{\mathbb Z}}$.

A (graded) pairing between DB cohomology spaces can be introduced. In our particular case of interest, it reduces to a commutative product:
\begin{equation}
\label{4}
\ast_D : H_D^1 \left( {M,{\mathbb Z}} \right) \times H_D^1 \left( {M,{\mathbb Z}}
\right)\longrightarrow H_D^{3} \left( {M,{\mathbb Z}}
\right)\simeq{{\mathbb R}/{\mathbb Z}} \, .
\end{equation}

\noindent The "DB square" of a class $ [ \omega ] \in H_D^1 \left( {M,{\mathbb Z}} \right)$ :
\begin{equation}
\label{5}
cs_1([ \omega ]) \equiv [ \omega ] \ast_D [ \omega ] \, .
\end{equation}
 canonically identifies with the abelian Chern-Simons (CS) lagrangian, while the level $k$  CS lagrangian simply reads
\begin{equation}
\label{6}
cs_k([ \omega ]) \equiv k \cdot cs_1([ \omega ]) = k \cdot [ \omega ] \ast_D [ \omega ] \, .
\end{equation}
Of course, due to the  $\mathbb Z$-module structure of DB spaces, $cs_k([ \omega ])$ is belonging to
$H_D^3 \left( {M,{\mathbb Z}} \right)$  if and only if  $k\in{\mathbb Z}$.

In fact, DB classes are another point of view for what is called Cheeger-Simons Differential Characters
(see for instance \cite{CS,Ko,Br,HLZ,BGST}). This implies that any DB cohomology class can be integrated
over any (integral) cycle of $M$ of the corresponding dimension. However, the result takes values in
${\mathbb R}/{\mathbb Z}$, and not ${\mathbb R}$ like in standard integration. Integral 3-cycles on an
oriented  3-manifold are just integer multiples of $M$. Hence, the lagrangian $cs_k([ \omega ])$ defines
the well known level $k$ CS action
\begin{equation}
\label{7}
CS_k([ \omega ]) \equiv k \int_M {cs_1([ \omega ])}
= k \int_M {[ \omega ] \ast_D [ \omega ]} \, .
\end{equation}
which takes values in ${\mathbb R}/{\mathbb Z}$ if and only if $k\in{\mathbb Z}$. We now have all the necessary ingredients to try to define the functional "CS measure" on $H_D^1 \left( {M,{\mathbb Z}} \right)$, denoted by
\begin{equation}
\label{8}
\mu _k \left( {\left[ \omega \right]} \right) \equiv D\left[ \omega \right] \cdot \exp\left\{
{2i\pi k\int_M  {\left[ \omega \right] \ast_D \left[ \omega \right]} } \right\} \, .
\end{equation}
Let us point out that (\ref{8}) imposes quantization of the level $k$, that is to say
\begin{equation}
\label{9}
k\in{\mathbb Z} \, ,
\end{equation}
for the exponential to be well defined. The procedure giving a meaning to (\ref{8}) was detailed in \cite{GT}.
To make it short let us say that, if we choose the exact sequence (\ref{1}) as defining
$H_D^1 \left( {M,{\mathbb Z}} \right)$, the measure will be made of a discrete sum indexed by elements
of $\check{H}^{2}\left( {M,{\mathbb Z}} \right)$; then, we pick up an origin on every (affine) fiber and for each of these fibers we consider a (formal) measure over the translation group ${\Omega ^1\left( M \right)} \mathord{\left/
{\vphantom {{\Omega ^1\left( M \right)} {\Omega _{\mathbb Z}^1 \left( M \right)}}} \right. \kern-\nulldelimiterspace} {\Omega _{\mathbb Z}^1 \left( M \right)}$. As already noted and extensively used in \cite{GT}, the CS measure satisfies
\begin{equation}
\label{10}
\mu _k \left( {\left[ \omega \right]} + \bar{\alpha}) \right) = \mu _k \left( {\left[ \omega \right]} \right ) \cdot
 \exp\left\{{2i\pi k\int_M \left( {2 \left[ \omega \right] \ast_D \bar{\alpha} + \bar{\alpha} \ast_D \bar{\alpha} } \right) } \right\} \, .
\end{equation}
for all $\bar{\alpha} \in {\Omega ^1\left( M \right)} \mathord{\left/
{\vphantom {{\Omega ^1\left( M \right)} {\Omega _{\mathbb Z}^1 \left( M \right)}}} \right. \kern-\nulldelimiterspace} {\Omega _{\mathbb Z}^1 \left( M \right)}$, which is similar to the Cameron-Martin property cylindrical functional measures verify.

In addition to the product $\ast_D$, integration of elements of $H_D^1 \left( {M,{\mathbb Z}} \right)$
over 1-cycles on $M$ is also providing a pairing:
\begin{equation}
\label{11}
\oint : H_D^1 \left( {M,{\mathbb Z}} \right) \times Z_1 \left( {M}
\right)\longrightarrow {{\mathbb R}/{\mathbb Z}} \, ,
\end{equation}
where $Z_1 \left( {M}\right)$ denotes the abelian group of (integral) 1-cycles on $M$. This pairing allows us to see 1-cycles
on $M$ as elements of $H_D^1 \left( {M,{\mathbb Z}} \right)^{\ast} \equiv Hom\left( {H_D^1 \left( {M,{\mathbb Z}} \right),{{\mathbb R}/{\mathbb Z}}} \right)$, the Pontrjagin dual
of $H_D^1 \left( {M,{\mathbb Z}} \right)$. This dual space is itself embedded into dual sequences (\cite{HLZ}):
\begin{equation}
\label{11b}
0\buildrel \over \longrightarrow \check{H}^{1}\left( {M,{\mathbb R
}/{\mathbb Z}}\right)\buildrel \over \longrightarrow H_D^1 \left( {M,{\mathbb Z}}
\right)^{\ast}\buildrel \over \longrightarrow Hom\left(  {\Omega ^1\left( M \right)} \mathord{\left/
{\vphantom {{\Omega ^1\left( M \right)} {\Omega _{\mathbb Z}^1 \left( M
\right)}}} \right. \kern-\nulldelimiterspace} {\Omega _{\mathbb Z}^1 \left( M
\right)},{{\mathbb R}/{\mathbb Z}} \right)\buildrel \over \longrightarrow 0 \, ,
\end{equation}
and
\begin{equation}
\label{12}
0\buildrel \over \longrightarrow Hom\left( \Omega _{\mathbb Z}^2 \left( M \right),{{\mathbb R}/{\mathbb Z}} \right)\buildrel
 \over \longrightarrow H_D^1 \left( {M,{\mathbb Z}}
\right)^{\ast}\buildrel \over \longrightarrow \check{H}^{2}\left( {M,{\mathbb Z}}
\right)\buildrel \over \longrightarrow 0 \, ,
\end{equation}
both being very similar to the original sequences (\ref{1}) and (\ref{2}).
On the other hand, the DB product (\ref{4}) is also allowing us to canonically identify $H_D^1 \left( {M,{\mathbb Z}}
\right)$ as a subspace of $H_D^1 \left( {M,{\mathbb Z}}\right)^{\ast}$ via integration over $M$, what is also legitimated by the  sequences above. But since  $Z_1 \left( {M}\right) \subset H_D^1 \left( {M,{\mathbb Z}}\right)^{\ast}$, one is naturally led to consider the possibility to associate to each  1-cycle,  $z$, on $M$ a (distributional) DB class, ${\left[ \eta_z \right]}$. Details of this association can be found in \cite{BGST}. These arguments look totally similar to how smooth functions can be considered as distributions via standard integration, and how chains can be seen as de Rham currents, except that everything is done
with respect to ${{\mathbb R}/{\mathbb Z}}$ and not ${\mathbb R}$.

The usefulness of the Pontrjagin dual in our problem is deeply related to the fact that, in Quantum Field Theory, the Quantum Configuration Space is made of distributional objects, and not just smooth ones. The first consequence will be an attempt to extend the CS measure to $H_D^1 \left( {M,{\mathbb Z}}\right)^{\ast}$ . However, while the DB product (\ref{4}) obviously extends to
\begin{equation}
\label{13}
\ast_D : H_D^1 \left( {M,{\mathbb Z}} \right) \times H_D^1 \left( {M,{\mathbb Z}}
\right)^{\ast}\longrightarrow {{\mathbb R}/{\mathbb Z}} \, ,
\end{equation}
it is hopeless to try to extend it straightforwardly to
\begin{equation}
\label{14}
\ast_D : H_D^1 \left( {M,{\mathbb Z}} \right)^{\ast} \times H_D^1 \left( {M,{\mathbb Z}}
\right)^{\ast} \longrightarrow {{\mathbb R}/{\mathbb Z}} \, ,
\end{equation}
since we will face the problem of defining product of distributions (or currents). Actually, we won't really need to give a meaning to the products of any two elements of $H_D^1 \left( {M,{\mathbb Z}} \right)^{\ast}$. We will only need to define products like  ${\left[ \eta_z \right]} \ast_D {\left[ \eta_z \right]}$, where ${\left[ \eta_z \right]}$  is the DB representative of a  1-cycle, $z$, on $M$. For the rest, we just need to assume that there is a functional measure on the Quantum Configuration Space ($\subseteq H_D^1 \left( {M,{\mathbb Z}} \right)^{\ast}$) which satisfies the Cameron-Martin like property (\ref{10}) (see \cite{GT} and references therein concerning this point).

Let us now deal with Wilson lines. We will explicitly consider $M=\mathbb{R}P^3$, although our treatment is quite oviously general.

\section{Expectation value of Wilson lines with torsion in the abelian Chern-Simons theory: $M=\mathbb{R}P^3$ case.}

The 3-manifold $M=\mathbb{R}P^3$ is among  the simplest ones involving torsion. Indeed, and due to Poincaré duality, we have:
\bea
\label{15}
\check{H}^{2}\left( {M,{\mathbb Z}} \right) \simeq \check{H}_{1}\left( {M,{\mathbb Z}} \right) = \mathbb{Z}_2 \\ \nonumber
\check{H}^{1}\left( {M,{\mathbb Z}} \right) \simeq \check{H}_{2}\left( {M,{\mathbb Z}} \right) = 0 \, .
\eea
The first equation, together with (\ref{1}), implies that $H_D^1 \left( {M,{\mathbb Z}} \right)$  is an affine fiber bundle with base space $\mathbb{Z}_2\equiv\{\check{0},\check{1}\}$, with $2\cdot\check{1}=\check{0}$. The fiber over $\check{0}$ clearly contains the zero $U(1)$-connection,  $\left[ 0 \right]$, which plays the role of a canonical origin in this fiber, so that a DB class $\left[ \omega_0 \right]$  over $\check{0}$ satisfies
\begin{equation}
\label{16}
\left[ \omega_0 \right] = \left[ 0 \right] + \bar{\alpha} \, ,
\end{equation}
for some $\bar{\alpha} \in {\Omega ^1\left( M \right)} \mathord{\left/
{\vphantom {{\Omega ^1\left( M \right)} {\Omega _{\mathbb Z}^1 \left( M \right)}}} \right. \kern-\nulldelimiterspace} {\Omega _{\mathbb Z}^1 \left( M \right)}$. Over $\check{1}$ there is unfortunately no such canonical choice. Nevertheless, from the exact sequence (\ref{12}), we see that $H_D^1 \left( {M,{\mathbb Z}} \right)^{\ast}$ is also an affine bundle with base space $\mathbb{Z}_2\equiv\{\check{0},\check{1}\}$. Thus, the choice of $\left[ 0 \right]$ for origin on the fiber over $\check{0}$ still holds. Now, as explained in [GT], and because of the inclusion $Z_1 \left( {M}\right) \subset H_D^1 \left( {M,{\mathbb Z}}\right)^{\ast}$, there is a family of "natural" choices of origin for the fiber over $\check{1}$ provided by  1-cycles,  $z$, on  $M$, or rather by their DB representatives ${\left[ \eta_z \right]}$. All we have to assume is that such an origin also belongs to the Quantum Configuration Space of the theory. We can then formally write the functional CS measure on $H_D^1 \left( {M,{\mathbb Z}} \right)^{\ast}$:

\newpage
\bea
\label{17}
\mu _k \left( {\left[ \omega \right]} \right) \equiv D\bar{\alpha} \cdot \exp\left\{
{2i\pi k\int_M  \bar{\alpha} \ast_D \bar{\alpha} } \right\} + \hspace{6cm} \\  \nonumber + D\bar{\alpha} \cdot \exp\left\{
{2i\pi k\int_M  ({\left[ \eta_1 \right]} + \bar{\alpha}) \ast_D ({\left[ \eta_1 \right]} + \bar{\alpha}) } \right\} \, .
\eea
where ${\left[ \eta_1 \right]}$ is the origin on the fiber over $\check{1}$ associated to some given (and so fixed) torsion cycle $\tau_1$ on $M$. In the second term of (\ref{17}) there appear the quantity ${\left[ \eta_1 \right]} \ast_D {\left[ \eta_1 \right]}$  which is ill defined as being a product of distributions (or rather de Rham currents). This is where regularization is required. Actually, and as mentioned earlier, regularization is only required later on when computing expectation values of Wilson lines. However, as we will see (check \cite{GT}), the quantities to regularize are of the type ${\left[ \eta_1 \right]} \ast_D {\left[ \eta_1 \right]}$. This is why we are going to deal with regularization right now.

\subsection{Regularization of ${\left[ \eta_1 \right]} \ast_D {\left[ \eta_1 \right]}$ via framing: linking numbers of torsion cycles.}

When a cycle $z$  is trivial, \textit{i.e.}  $z=bc$ with $b$  the usual boundary operator, one can define the self linking number of $z$ as the linking number of $z$ with $z^f$, where $z^f$ is a framing of $z$. This reads:
\begin{equation}
\label{18}
L(z,z) \equiv L(z,z^f) \equiv c \ti z^f \, ,
\end{equation}
with $\ti$ denoting the transverse intersection. Of course, the result fully depends on the chosen framing of $z$. This also provides a regularization procedure for ${\left[ \eta_1 \right]} \ast_D {\left[ \eta_1 \right]}$. Indeed, if  $z$ and $z'$ are two trivial cycles in $M$ without any common points, their DB representatives, ${\left[ \eta_z \right]}$ and ${\left[ \eta_{z'} \right]}$, satisfy
\begin{equation}
\label{19}
{\left[ \eta_z \right]} \ast_D {\left[ \eta_{z'} \right]} = \left[ 0 \right] + \overline{\eta_z \wedge d\eta_{z'}} \in H_D^3 \left( {M,{\mathbb Z}}\right)^{\ast} \equiv {{\mathbb R}/{\mathbb Z}} \, ,
\end{equation}
where $\eta_z$ (resp. $\eta_{z'}$) is the de Rham current of the cycle $z$ (resp. $z'$) such that $z=bc$ (resp. $z'=bc'$). But $\eta_z \wedge d\eta_{z'}$ is the de Rham current representing the intersection $c \ti z' = c' \ti z$. Accordingly, $\int_M \eta_z \wedge d\eta_{z'} \in \mathbb{Z}$, so that ${\left[ \eta_z \right]} \ast_D {\left[ \eta_{z'} \right]} = \left[ 0 \right]$. Note that we didn't use any regularizing at this stage. We can now apply this to $z$ and $z^f$, leading to ${\left[ \eta_z \right]} \ast_D {\left[ \eta_{z^f} \right]}= \left[ 0 \right]$. Thus, the framing procedure can be used to regularize ${\left[ \eta_z \right]} \ast_D {\left[ \eta_z \right]} $ into $\left[ 0 \right]$. It can even be applied for a non trivial (but torsionless) cycle (see \cite{BGST,GT} for details).

For two torsion cycles $\tau$ and $\tau'$  on $M$  we have $2\tau = b \zeta$ and $2\tau' = b \zeta'$. Hence,  $\zeta \ti \tau'$ and $\zeta' \ti \tau$ are still well defined integers. The linking number of these torsion cycle is then
\begin{equation}
\label{20}
L(\tau,\tau') = \frac{1}{2} \, \zeta \ti \tau' \in  \frac{1}{2} \, \mathbb{Z} \, .
\end{equation}
Due to the occurrence of one half factor in (\ref{20}), we immediately conclude that there is no chance for the framing procedure to regularize ${\left[ \eta_{\tau} \right]} \ast_D {\left[ \eta_{\tau} \right]}$ into $\left[ 0 \right]$. Accordingly, the term ${\left[ \eta_1 \right]} \ast_D {\left[ \eta_1 \right]}$ appearing within (\ref{17}) will plague the CS measure since, by construction, it is built from a torsion cycle. Fortunately, there is the level parameter $k$ also occurring in (\ref{17}). Now, if $k=2l$, then $k{\left[ \eta_1 \right]} \ast_D {\left[ \eta_1 \right]} = l \cdot 2{\left[ \eta_1 \right]} \ast_D {\left[ \eta_1 \right]}$, and hence the framing procedure consistently applies to $2{\left[ \eta_1 \right]} \ast_D {\left[ \eta_1 \right]}$ because the factor one half into (\ref{20}) is now vanishing. Thus, here comes a new constraint on the CS level parameter for $M=\mathbb{R}P^3$:
\begin{equation}
\label{21}
k = 2l, \, l \in  \mathbb{Z} \, .
\end{equation}
Note that one could decide to regularize by using only an "even" framing, keeping $k \in  \mathbb{Z}$. But obviously this would be totally equivalent to consider any framing and $k = 2l$. This is this last point of view we will chose and from now on $k$ will be even.

We are now ready to look at Wilson lines.

\subsection{Expectation value of a Wilson line on $M=\mathbb{R}P^3$: trivial cycles and torsion cycles with charge $q$.}

Let $z$ be a 1-cycle on $M=\mathbb{R}P^3$. As previously explained, for any $\left[ \omega \right] \in H_D^1 \left( {M,{\mathbb Z}}\right)$
\begin{equation}
\label{22}
\int_z \left[ \omega \right] \in {{\mathbb R}/{\mathbb Z}} \, .
\end{equation}
This integral defines parallel transport of the connection $\left[ \omega \right]$  along the cycle $z$, and
\begin{equation}
\label{23}
\exp \left\{ 2i \pi \int_z \left[ \omega \right] \right\} \, .
\end{equation}
is called the $U(1)$-holonomy of $z$ with respect to the connection (or to the DB class) $\left[ \omega \right]$. We also noticed that it is possible to write
\begin{equation}
\label{24}
\int_z \left[ \omega \right] \mathop  = \limits_\mathbb{Z}  \int_M \left[ \omega \right] \ast_D {\left[ \eta_z \right]} \, .
\end{equation}
for ${\left[ \eta_z \right]} \in H_D^1 \left( {M,{\mathbb Z}} \right)^{\ast}$ canonically representing $z$. As long as $\left[ \omega \right]$ is smooth, formula (\ref{24}) is well defined, but since we need to go to $H_D^1 \left( {M,{\mathbb Z}} \right)^{\ast}$, once more some regularization will be required. On the other hand, a fundamental loop is a continuous mapping,  $f: \rightarrow S^1 M$, such that $f(S^1) \simeq S^1$. A singular decomposition of $S^1$ provides a singular decomposition of $f(S^1)$ so that this last quantity can be considered as a (singular) 1-cycle on $M$. Then, we can consider linear combinations:
\begin{equation}
\label{25}
z = \sum_{i}^{N} q_i Z_i \, .
\end{equation}
where the $Z_i$ are fundamental loops without any common points.

>From now on, we will assume that the functional CS measure is (existing and) normalized so that:
\begin{equation}
\label{26}
\int \mu _k \left( {\left[ \omega \right]} \right) = 1 \, .
\end{equation}
The expectation values of the Wilson line for a fundamental loop $Z$ with respect to the level $k$ CS measure formally reads
\begin{equation}
\label{27}
\left\langle W(Z) \right\rangle_k
\equiv \left\langle \exp \left\{ 2i \pi \int_Z \left[ \omega \right] \right\} \right\rangle
\equiv \int \mu _k \left( {\left[ \omega \right]} \right) \exp \left\{ 2i \pi \int_Z \left[ \omega \right] \right\} \, ,
\end{equation}
and for cycle a $z=qZ$
\begin{equation}
\label{28}
\left\langle W(z=qZ) \right\rangle_k
= \int \mu _k \left( {\left[ \omega \right]} \right) \exp \left\{ 2iq \pi \int_Z \left[ \omega \right] \right\} \, .
\end{equation}
>From (\ref{24}) we can equivalently write
\begin{equation}
\label{29}
\left\langle W(z=qZ) \right\rangle_k
= \int \mu _k \left( {\left[ \omega \right]} \right)
\exp \left\{ 2i \pi q \int_M \left[ \omega \right] \ast_D \left[ \eta_Z \right] \right\} \, .
\end{equation}
finally injecting (\ref{17}) into (\ref{29}) we obtain
\bea
\label{30}
\left\langle W(z=qZ) \right\rangle_k
&& = \int D\bar{\alpha}
\exp\left\{{2i\pi \int_M  \bar{\alpha} \ast_D (k\bar{\alpha} + q\left[ \eta_Z \right]) } \right\}
 \\ \nonumber
&& + \int D\bar{\alpha} \exp\left\{{2i\pi \int_M  ({\left[ \eta_1 \right]} + \bar{\alpha})
\ast_D (k{\left[ \eta_1 \right]} + k\bar{\alpha} + q\left[ \eta_Z \right]) } \right\}  \, .
\eea
There are two different cases to consider: either $Z=bC$ (trivial cycle), or $2Z=bC'$ but $Z \neq bC$ (torsion cycle).

When $Z=bC$ and with our choice of origin on the trivial fiber of $H_D^1 \left( {M,{\mathbb Z}} \right)^{\ast}$, we can write $\left[ \eta_Z \right] = \overline{\beta_C}$ for some $\overline{\beta_C} \in Hom\left( \Omega _{\mathbb Z}^2 \left( M \right),{{\mathbb R}/{\mathbb Z}} \right)$.  As explained in \cite{BGST}, $\overline{\beta_C}$ is built from the de Rham current,  $\beta_C$, of the chain $C$. Unlike DB classes,   $\beta_C$ can be divided by $2k$  giving rise to $\overline {{{\beta _C } \mathord{\left/ {\vphantom {{\beta _C } {2k}}} \right. \kern-\nulldelimiterspace} {2k}}} \in Hom\left( \Omega _{\mathbb Z}^2 \left( M \right),{{\mathbb R}/{\mathbb Z}} \right)$. Now, as intensively done in \cite{GT}, we perform the shift
\begin{equation}
\label{31}
\bar{\alpha} \rightarrow \bar{\chi} = \bar{\alpha}
+ q \overline{\frac{\beta _C}{2k}}
\, .
\end{equation}
in both terms of (\ref{30}), thus obtaining
\bea
\label{32}
\left\langle W(z=qZ) \right\rangle_k
= \int D\bar{\chi}
\exp\left\{{2i\pi k \int_M  \bar{\chi} \ast_D \bar{\chi} } \right\}
\exp\left\{{-2i\pi kq^2 \int_M  \overline{\frac{\beta _C}{2k}} \ast_D \overline{\frac{\beta _C}{2k}} } \right\}
 \\ \nonumber
 + \int D\bar{\chi} \exp\left\{{2i\pi k \int_M  ({\left[ \eta_1 \right]} + \bar{\chi})
\ast_D ({\left[ \eta_1 \right]} + \bar{\chi}) } \right\}
\exp\left\{{-2i\pi kq^2 \int_M  \overline{\frac{\beta _C}{2k}} \ast_D \overline{\frac{\beta _C}{2k}} } \right\}  \, ,
\eea
where we used: $2k\overline {{{\beta _C } \mathord{\left/ {\vphantom {{\beta _C } {2k}}} \right. \kern-\nulldelimiterspace} {2k}}}
\mathop  = \limits_\mathbb{Z} \overline{\beta_C}$. Note that the result mainly derives from the Cameron-Martin property of the CS measure. Finally, since
\begin{equation}
\label{33}
\overline{\frac{\beta _C}{2k}} \ast_D \overline{\frac{\beta _C}{2k}} \mathop  = \limits_\mathbb{Z}
\overline{\frac{\beta _C}{2k} \wedge d\frac{\beta _C}{2k}}
\mathop  = \limits_\mathbb{Z} \overline{\frac{\beta _C \wedge d\beta _C}{4k^2}}
\, ,
\end{equation}
we derive
\begin{equation}
\label{34}
\exp\left\{{-2i\pi kq^2 \int_M  \overline{\frac{\beta _C}{2k}} \ast_D \overline{\frac{\beta _C}{2k}} } \right\}
 = \exp\left\{{-\frac{2i\pi q^2}{4k} \int_M \beta _C \wedge d\beta _C } \right\} \, .
\end{equation}
The product $\beta _C \wedge d\beta _C$  has to be regularized for its integral over $M$  to have a meaning. Applying the framing procedure to  $Z$, leads to
\begin{equation}
\label{35}
\int_M \beta _C \wedge d\beta _C  \equiv L(Z,Z^f) \equiv C \ti Z^f \in \mathbb{Z} \, .
\end{equation}
We then conclude that
\begin{equation}
\label{36}
\left\langle W(z=qZ) \right\rangle_k
= \exp\left\{{-2i\pi \frac{q^2}{4k} L(Z,Z^f)} \right\} = \exp\left\{{-2i\pi \frac{q^2}{4k} C \ti Z^f } \right\}
 \, ,
\end{equation}
which is, as expected, the same result as for $M=S^3$. Let us prove that the above procedure doesn't depend on our choice of $\beta _C$. Let $\tilde{C}$ be another chain bounding  $Z$. Then $b(\tilde{C} - C) = 0$  which means that $\tilde{C} - C$  is a 2-cycle on  $M$. Since here $M=\mathbb{R}P^3$, from (\ref{15}) we deduce that $\tilde{C} - C = b\vartheta$. Then $b\vartheta \ti Z^f = \vartheta \ti bZ^f = 0$, and (\ref{36}) will still hold. If $M$ has free homology of degree two, there will also be free cohomology of degree two (see Universal coefficient theorem), and then the base space of $H_D^1 \left( {M,{\mathbb Z}} \right)$  (and  $H_D^1 \left( {M,{\mathbb Z}} \right)^{\ast}$) will also have a free part so that we have to adapt our measure. However, it is almost obvious that (\ref{34}) would then produce a term  $(\tilde{C} - C) \ti Z^f = (\tilde{C} - C) \ti bC^f = b(\tilde{C} - C) \ti C^f = 0$, since by hypothesis $Z$, and so $Z^f$, is a trivial cycle.

In the torsion case, since $2Z=bC'$, we can obviously write $\left[ \eta_{2Z} \right] = 2\left[ \eta_{Z} \right] = \overline{\beta_{C'}}$, with $\overline{\beta_{C'}}$ built from the de Rham current,  $\beta_{C'}$, of the chain $C'$. However, since  $Z\neq bC$, we cannot find any de Rham current $\beta_{C}$ of an integral chain such that $\left[ \eta_{Z} \right] = \overline{\beta_{C}}$. This is because DB cohomology is defined over $\mathbb{Z}$ and not $\mathbb{Q}$. On the other hand, $\left[ \eta_1 \right]$, the DB representative of the fixed torsion cycle $\tau_1$, has been chosen as origin of the fiber over $\check{1}$, so we can also write $\left[ \eta_Z \right] = \left[ \eta_1 \right] + \overline{\beta_y}$, where $\overline{\beta_y}$ is  made from the de Rham current, $\overline{\beta_y}$, of the chain $y$ relating $Z$ and $\tau_1$: $Z = \tau_1 + by$. Injecting that into (\ref{30}) gives
\bea
\label{37}
\left\langle W(z=qZ) \right\rangle_k \hspace{-0.5cm}
&& = \int D\bar{\alpha}
\exp\left\{{2i\pi \int_M  \bar{\alpha} \ast_D (k\bar{\alpha} + q\left[ \eta_1 \right] + q\overline{\beta_y}) } \right\}
 \\ \nonumber
&& + \int D\bar{\alpha} \exp\left\{{2i\pi \int_M  ({\left[ \eta_1 \right]} + \bar{\alpha})
\ast_D (k{\left[ \eta_1 \right]} + k\bar{\alpha} + q\left[ \eta_1 \right]) + q\overline{\beta_y}} \right\}  \, .
\eea
Since $k$ is even, the quantity $k{\left[ \eta_1 \right]} \ast_D {\left[ \eta_1 \right]}$ occurring in the second term of this expression is consistently regularized into $\left[ 0 \right]$ using the framing procedure. Unfortunately, in the same term we also see the quantity $q{\left[ \eta_1 \right]} \ast_D {\left[ \eta_1 \right]}$. It combines with the previous one to give $(k+q){\left[ \eta_1 \right]} \ast_D {\left[ \eta_1 \right]}$. From the same regularization argument which led us to impose $k$ to be even, we deduce that $(k+q)$  has to be even too, and thus
\begin{equation}
\label{38}
q = 2m, \, m \in  \mathbb{Z} \, .
\end{equation}
In other words, charges inherit the same constraint than the level parameter and for exactly the same reasons. Note that when $q$  is odd then the framing procedure might produce variations of the relative sign between the two terms of (\ref{37}), depending whether the framing is odd or even, hence implying that the expectation value wouldn't be properly defined.
Let us assume for the rest of this section that $q=2m$, and let us rewrite (\ref{37}) accordingly:
\bea
\label{39}
\left\langle W(z=qZ) \right\rangle_k \hspace{-0.5cm}
&& = \int D\bar{\alpha}
\exp\left\{{2i\pi \int_M  \bar{\alpha} \ast_D (k\bar{\alpha} + 2m\left[ \eta_1 \right] + 2m\overline{\beta_y}) } \right\}
 \\ \nonumber
&& + \int D\bar{\alpha} \exp\left\{{2i\pi \int_M  ({\left[ \eta_1 \right]} + \bar{\alpha})
\ast_D (k{\left[ \eta_1 \right]} + k\bar{\alpha} + 2m\left[ \eta_1 \right]) + 2m\overline{\beta_y}} \right\}  \, .
\eea
Since $\left[ \eta_1 \right]$ is the DB representative of the torsion cycle $\tau_1$, there exist a chain $C$  with de Rham current $\gamma_C$, such that $2\tau = bC$, that is to say $2\left[ \eta_1 \right] = \overline{\gamma_C}$. Hence
\bea
\label{40}
\left\langle W(z=qZ) \right\rangle_k \hspace{-0.2cm}
&=& \int D\bar{\alpha}
\exp\left\{{2i\pi \int_M  \bar{\alpha} \ast_D (k\bar{\alpha} + m\overline{\gamma_C} + 2m\overline{\beta_y}) } \right\}
 \\ \nonumber
 + && \int D\bar{\alpha} \exp\left\{{2i\pi \int_M  ({\left[ \eta_1 \right]} + \bar{\alpha})
\ast_D (k{\left[ \eta_1 \right]} + k\bar{\alpha} + m\overline{\gamma_C} + 2m\overline{\beta_y} ) } \right\} \\ \nonumber
&=& \int D\bar{\alpha}
\exp\left\{{2i\pi \int_M  \bar{\alpha} \ast_D (k\bar{\alpha} + m \overline{\rho_{C+2y}} ) } \right\}
 \\ \nonumber
 + && \int D\bar{\alpha} \exp\left\{{2i\pi \int_M  ({\left[ \eta_1 \right]} + \bar{\alpha})
\ast_D (k{\left[ \eta_1 \right]} + k\bar{\alpha} + m\overline{\rho_{C+2y}} ) } \right\}
 \, .
\eea
where we have introduce $\overline{\rho_{C+2y}} = \overline{\gamma_C} + 2\overline{\beta_y} = \overline{\gamma_C + 2\beta_y}$, with $\rho_{C+2y}$ being the de Rham current of $C+2y$. Now, let us perform the usual shift
\begin{equation}
\label{42}
\bar{\alpha} \rightarrow \bar{\chi} = \bar{\alpha}
+ q \overline{\frac{\rho_{C+2y}}{2k}}
\, .
\end{equation}
to obtain
\bea
\label{43}
\left\langle W(z=qZ) \right\rangle_k
= \int D\bar{\chi}
\exp\left\{{2i\pi k \int_M  \bar{\chi} \ast_D \bar{\chi} } \right\}
\exp\left\{{-2i\pi km^2 \int_M  \overline{\frac{\rho_{C+2y}}{2k}} \ast_D \overline{\frac{\rho_{C+2y}}{2k}} } \right\}
 \\ \nonumber
 + \int D\bar{\chi} \exp\left\{{2i\pi k \int_M  ({\left[ \eta_1 \right]} + \bar{\chi})
\ast_D ({\left[ \eta_1 \right]} + \bar{\chi}) } \right\}
\exp\left\{{-2i\pi km^2 \int_M  \overline{\frac{\rho_{C+2y}}{2k}} \ast_D \overline{\frac{\rho_{C+2y}}{2k}} } \right\}  \, .
\eea
We are left with proving that the framing procedure is providing a consistent regularization of $\overline {{{\rho_{C+2y}} \mathord{\left/ {\vphantom {{\rho _C } {2k}}} \right. \kern-\nulldelimiterspace} {2k}}} \ast_D \overline {{{\rho_{C+2y}} \mathord{\left/ {\vphantom {{\rho _C } {2k}}} \right. \kern-\nulldelimiterspace} {2k}}}$, giving (\ref{42}) a meaning. Actually, if $Z^f$ denotes a framing of $Z$:
\begin{equation}
\label{44}
km^2 \int_M  \overline{\frac{\rho_{C+2y}}{2k}} \ast_D \overline{\frac{\rho_{C+2y}}{2k}}
\mathop  = \limits_\mathbb{Z} \frac{m^2}{4k} \int_M  \rho_{C+2y} \wedge d\rho_{C+2y}
=  \frac{m^2}{4k} \cdot (C+2y) \ti 2Z^f \, ,
\end{equation}
what implies
\bea
\label{43}
\left\langle W(z=qZ) \right\rangle_k = exp\left\{-2i \pi \frac{q^2}{4k} \cdot \frac{(C+2y) \ti Z^f}{2} \right\} \, .
\eea
We also introduce the 2-chain $C'$ such that $2Z = bC'$. Hence, $b(C' - C - 2y) = 0$, which means that $C' - C - 2y$ is a 2-cycle on $M$. Since the second homology group of $\mathbb{R}P^3$ is trivial, in this case $C' - C - 2y = b\vartheta$ what implies $(C+2y) \ti Z^f = C' \ti Z^f$. Once more, if $M$ had a non trivial second homology group, then we would have $(C+2y) \ti Z^f = C' \ti Z^f + \Sigma \ti Z^f$ for some (possibily non trivial) 2-cycle. Yet, since $2Z^f=bCf$ we would still obtain that $(C+2y) \ti Z^f = C' \ti Z^f$.
Finally
\bea
\label{44}
\left\langle W(z=qZ) \right\rangle_k = exp\left\{-2i \pi  \frac{q^2}{4k} \cdot \frac{C' \ti Z^f}{2} \right\} \, ,
\eea
with $2Z=bC'$, which is exactly the result coming from surgery \cite{Ro,Gu,Mo}. This last series of results also prove that nothing depends on the choice we made for $\rho_{C+2y}$.

Finally, note that (\ref{44}) is actually containing (\ref{36}) since if $2Z=bC'$ and $Z=bC$ then $C'=2C$ is a possible choice and then $C' \ti Z^f / 2 = C \ti Z^f$ has expected. And consistently, we don't need  $q$ to be even within (\ref{36}). One can convince himself that the factor 2 appearing in (\ref{43}) is nothing but the torsion degree of $Z$, and thus in the case of a 3-manifold with torsion cycle of degree $p$ we would see a term like $C' \ti Z^f / p = C \ti Z^f$. This is also in agreement with the case of trivial cycles which can be seen as torsion cycles of degree 1.

\section{Conclusions}

The treatment of abelian Chern-Simons to generate link invariants introduced in \cite{GT} straightforwardly extends to the case of oriented 3-manifolds with torsion. And although we only considered $\mathbb{R}P^3$, it is clear that our results apply to any oriented 3-manifold with torsion. In \cite{PT}, we will show how Deligne-Beilinson cohomology can also be applied to higher dimensional abelian Chern-Simons theories and links invariants, thus fulfilling some of the questions left opened in \cite{GT}.

\vskip 1 truecm

{\bf{Acknowledgments.}} I wish to thank Enore Guadagnini for many fruitful discussions in particular those concerning surgery related to link invariants involving torsion cycles on $\mathbb{R}P^3$.

\vfill\eject

\end{document}